\documentclass[aps,prc,10pt,twocolumn,superscriptaddress,showpacs]{revtex4-1}
\usepackage{graphicx}
\usepackage{color}

\begin{document}

\title{Special quasirandom structure in heterovalent ionic systems}
\author{Atsuto \surname{Seko}}
\email{seko@cms.mtl.kyoto-u.ac.jp}
\affiliation{Department of Materials Science and Engineering, Kyoto University, Kyoto 606-8501, Japan}
\affiliation{Center for Elements Strategy Initiative for Structure Materials (ESISM), Kyoto University, Kyoto 606-8501, Japan}
\author{Isao \surname{Tanaka}}
\affiliation{Department of Materials Science and Engineering, Kyoto University, Kyoto 606-8501, Japan}
\affiliation{Center for Elements Strategy Initiative for Structure Materials (ESISM), Kyoto University, Kyoto 606-8501, Japan}
\affiliation{Nanostructures Research Laboratory, Japan Fine Ceramics Center, Nagoya 456-8587, Japan}

\date{\today}


\begin{abstract}
The use of a special quasirandom structure (SQS) is a rational and efficient way to approximate random alloys.
A wide variety of physical properties of metallic and semiconductor random alloys have been successfully estimated by a combination of an SQS and density functional theory (DFT) calculation.
Here, we investigate the application of an SQS to the ionic multicomponent systems with configurations of heterovalent ions, including point-charge lattices, MgAl$_2$O$_4$ and ZnSnP$_2$.
It is found that the physical properties do not converge with the supercell size of the SQS.
This is ascribed to the fact that the correlation functions of long-range clusters larger than the period of the supercell are not optimized in the SQS. 
However, we demonstrate that the physical properties of the perfectly disordered structure can be estimated by linear extrapolation using the inverse of the supercell size. 
\end{abstract}

\maketitle

\section{Introduction}
\label{sqs:introduction}
\subsection{Special quasirandom structure (SQS)}
Density functional theory (DFT) calculation\cite{PhysRev.136.B864,PhysRev.140.A1133} enables us to compute a wide variety of physical properties. 
However, it is not directly applicable to random alloys since they are expressed only by average occupancies of constituent atoms.
The use of a special quasirandom structure (SQS)\cite{SQS} is one approach to approximating random alloys. 
In this approach, a periodic ordered structure having a similar atomic configuration to the perfectly disordered structure is used as mentioned below. 
By replacing the perfectly disordered structure with a periodic ordered structure, physical properties can be easily computed by DFT calculation.
Owing to the convenience, a combination of DFT calculation and an SQS has been widely used to estimate the physical properties of random alloys such as 
the mixing enthalpy\cite{PhysRevB.69.214202,PhysRevB.74.024204,olsson2006electronic,PhysRevB.75.045123,PhysRevB.42.9622,PhysRevB.76.144204,zhu2008electronic,PhysRevB.78.195201,PhysRevB.73.235214,PhysRevB.72.184203,sanati2003ordering,wolverton2006first,PhysRevB.80.174101},
lattice parameters\cite{PhysRevB.69.214202,PhysRevB.74.024204,olsson2006electronic,PhysRevB.75.045123},
elastic properties\cite{PhysRevB.81.094203},
magnetic properties\cite{PhysRevB.69.214202,PhysRevLett.94.097201},
paramagnetism\cite{PhysRevB.85.125104,Ikeda_sqs_paraFe},
lattice vibrational properties\cite{PhysRevB.73.235214,graanas2012new},
electronic structure and related properties such as the band gap\cite{Yokoyama_APEX.6.061201,PhysRevB.42.9622,PhysRevB.42.3757,apl/94/4/10.1063/1.3074499,PhysRevB.44.3387,PhysRevLett.80.4939,wang2007effects,wei1996giant,wei1998effects,wei2000first,bellaiche1997band,PhysRevB.77.033103,PhysRevB.87.241201,lee2011prediction,ZnSnP2_sqs_Walsh},
optical absorption spectrum\cite{Yokoyama_APEX.6.061201} 
and piezoelectric properties\cite{PhysRevLett.104.137601}.

The idea of the SQS was derived from the cluster expansion (CE) method\cite{CE1,CE2,CE3}.
Within the formalism of the CE method for binary alloys, a physical property $\Lambda$ for an alloy configuration is written as
\begin{equation}
\Lambda = \sum\limits_{\alpha} V_\alpha \cdot \varphi_\alpha , 
\label{hamiltonian}
\end{equation}
where $V_\alpha$ and $ \varphi_\alpha $ are called the effective cluster interaction (ECI) and the correlation function of cluster $\alpha$, which ranges from $-1$ to $+1$, respectively.
The correlation function is used to find an SQS similar to the perfectly disordered structure.
The similarity of two alloy configurations is usually measured by the squared norm of the difference of the correlation functions.
The similarity of a candidate structure and the perfectly disordered structure is expressed as 
\begin{equation}
\sum_\alpha \left| \varphi_\alpha - \varphi_\alpha^{\rm (disorder)}\right|^2,
\label{sqs:sqs-criterion}
\end{equation}
where $\varphi_\alpha^{\rm (disorder)}$ denotes the correlation function of cluster $\alpha$ for the perfectly disordered structure that is analytically given according to its composition.

\subsection{Cluster truncation for SQS}
The definition of similarity given by Eq. (\ref{sqs:sqs-criterion}) can be composed of an infinite number of clusters.
However, a small number of truncated pairs has been practically adopted to find the SQS.
For instance, in the original paper by Zunger $et$ $al$.\cite{SQS}, SQSs were obtained by minimizing a similarity defined by pairs up to the fourth nearest neighbor (NN) with a constraint that the correlation function of the first NN pair is exactly the same as that of the perfectly disordered structure.
Generally, the similarity must be defined without any a priori knowledge of ECIs.
Therefore, the cluster truncation has been determined from an empirical consideration or occasionally from a convergence test of the physical properties with increasing number of pairs.

In the general case of optimizing the correlation functions of only a small number of pairs, an SQS with a small number of atoms can be adopted.
Such an SQS can be explored exhaustively by calculating the correlation functions of all candidate alloy configurations expressed by a small number of atoms, which are obtained by a search for derivative structures\cite{Hart_derivativestructure,Hart_derivativestructure2}.
Among the candidates, the alloy configuration with the closest correlation functions to those of the perfectly disordered structure is regarded as the best SQS.

\begin{figure}[tbp]
\begin{center}
\includegraphics[width=\linewidth]{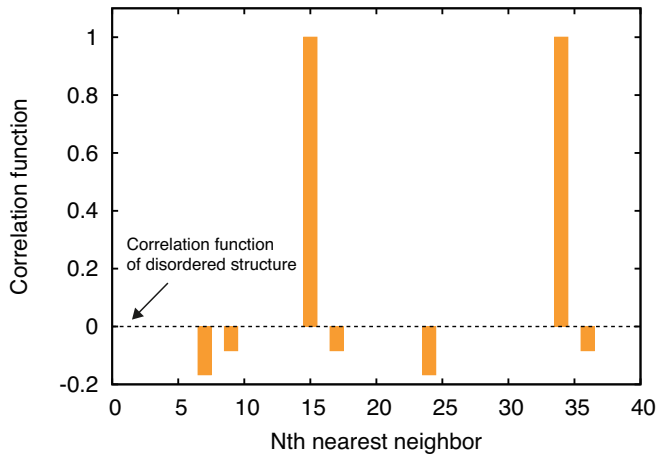} 
\caption{
Correlation functions of SQS with the composition of 0.5 in binary systems with the simple cubic lattice. 
The analytical correlation functions of the perfectly disordered structure are exactly zero.
The SQS is constructed by optimizing the correlation functions of pairs up to the fifth NN within $4 \times 4 \times 4$ supercell.
The correlation functions of the 15th and 34th NN pairs are exactly one because the 15th and 34th NN pairs are always composed of the same types of atoms owing to the periodicity of the supercell.
}
\label{sqs:sqs-correlation-sc}
\end{center}
\end{figure}

To guarantee the accuracy of the SQS, however, the numbers of atoms and clusters used for the SQS should be determined very carefully.
Let us consider an SQS with a composition of 0.5 in a binary system with the simple cubic lattice.
This SQS is constructed by optimizing the correlation functions of pairs up to the fifth NN by simulated annealing\cite{SA1,SA2} within the $4 \times 4 \times 4$ supercell of the simple cubic lattice.
The correlation functions of pairs up to 40th NN of the SQS are shown in Fig. \ref{sqs:sqs-correlation-sc}.
As can be seen in Fig. \ref{sqs:sqs-correlation-sc}, the correlation functions of several pairs deviate from those of the perfectly disordered structure.
In particular, the correlation functions of the 15th NN and 34th NN pairs are exactly unity owing to the periodicity of the supercell.
When the ECIs of these pairs are not negligible, the deviation of the correlation functions generates systematic errors.
To improve the accuracy of the SQS, it is necessary to increase the number of atoms of the SQS and/or the number of pair clusters used for the optimization.

With the exception of multicomponent metallic or isovalent ionic systems, the contributions of long-range ECIs to configurational energetics are not negligible in heterovalent ionic systems, which is ascribed to long-range electrostatic interactions\cite{Seko_longrange_Jphys}.
Therefore, careful validation of the convergence of long-range interactions is necessary to find an SQS in such systems.
In this study, we examine the error resulting from the use of an SQS for the calculation of physical properties in heterovalent ionic systems.
The dependence of the SQS energy on the numbers of atoms and pairs used to optimize the SQS will be demonstrated.
As heterovalent ionic systems, model systems described only with point charges on the spinel and zincblende lattices are first examined, hereafter called the ``point-charge spinel lattice" and ``point-charge zincblende lattice'', respectively.
The use of such model systems makes it easier to discuss the error of the SQS energy because the exact energy of the perfectly disordered structure can be easily computed.
Then, the SQS is applied to the DFT calculation of the energy, volume and band gap in actual MgAl$_2$O$_4$ and ZnSnP$_2$ systems.

\section{Energy of disordered structure in point-charge lattice systems}
\label{sqs:point-charge}
\subsection{Point-charge spinel and zincblende lattices}
To begin with, we consider cation disordering on a point-charge spinel lattice with the formula AB$_2$C$_4$ and a point-charge zincblende lattice with the formula ABC$_2$, where A and B are cations and C is an anion.
The spinel structure has two types of cation site, namely tetrahedral fourfold-coordinated and octahedral sixfold-coordinated sites. 
The number of octahedral sites is double the number of tetrahedral sites.
When all the tetrahedral sites are occupied by cation A, the spinel is called ``normal''. 
When all the tetrahedral sites are occupied by cation B, the spinel is called ``inverse''. 
In the zincblende lattice, cations occupy half of the fourfold-coordinated tetrahedral sites and form the face-centered cubic (fcc) lattice.
When cations A and B have the D0$_{22}$ configuration, the configuration is called the chalcopyrite structure.

In point-charge lattices, interatomic interactions are described only by the electrostatic interactions between point charges.
The electrostatic energy $E_{\rm es}$ for a point-charge configuration is expressed by
\begin{equation}
E_{\rm es} = \frac{1}{2} \sum_{i,j} \frac{q_i q_j}{r_{ij}},
\end{equation}
where $q_i$ and $r_{ij}$ denote the charge of lattice site $i$ and the distance between lattice sites $i$ and $j$, respectively.
We adopt a point-charge spinel lattice with $q_{\rm A} = +2$, $q_{\rm B} = +3$ and $q_{\rm C} = -2$, where only the cation disordering is considered.
The unit-cell shape is kept cubic.
The lattice constant and internal parameter are fixed to 8 \AA\ and 0.3855, respectively.
For the point-charge zincblende lattice, charges of $q_{\rm A} = +2$, $q_{\rm B} = +4$ and $q_{\rm C} = -3$ are used.
The lattice constant of the point-charge zincblende lattice is set to 5.5 \AA.
The electrostatic energy is evaluated by the Ewald method\cite{Ewald,Kittel:ISSP,Ziman:Principles} using the {\sc clupan} code\cite{clupan,ame_feature:Seko,casp:seko}.

The electrostatic energy of the perfectly disordered structure can be exactly evaluated using the average point charges of cation and anion sites, $q_{\rm cation}$ and $q_{\rm anion}$.
In the perfectly disordered structure of the point-charge spinel lattice, $q_{\rm cation} = +8/3$ and $q_{\rm anion} = -2$.
Similarly, for the point-charge zincblende lattice, the average point charges are $q_{\rm cation} = +3$ and $q_{\rm anion} = -3$.
The exact electrostatic energies of the perfectly disordered structure for the point-charge spinel and zincblende lattices are $-78.206$ and $-89.139$ eV/cation, respectively, as listed in Table \ref{sqs:point-charge-disorder-energy}.

\begingroup
\squeezetable
\begin{table}[tbp]
\caption{
Exact electrostatic energy of the perfectly disordered structure for the point-charge spinel and zincblende lattices (Unit: eV/cation).
Energies extrapolated from average Monte Carlo (MC) energies and SQS energies are also shown.
}
\label{sqs:point-charge-disorder-energy}
\begin{ruledtabular}
\begin{tabular}{lcccc}
& Exact & MC & \multicolumn{2}{c}{SQS}\\
\hline
Spinel & -78.206 & -78.215 & -78.194 (60th NN) & -78.249 (300th NN)\\
Zincblende & -89.139 & -89.176 & -89.097 (10th NN) & -89.118 (50th NN)
\end{tabular}
\end{ruledtabular}
\end{table}
\endgroup

\subsection{Supercell approaches}
To estimate the energy of the perfectly disordered structure, we adopt two types of supercell approach: Monte Carlo (MC) simulations\cite{MC1,MC2,MC3} at an extremely high temperature and SQS calculations.
In the MC simulations, several types of supercell are constructed by isotropic expansions of the primitive cell up to $8 \times 8 \times 8$ (7162 atoms) and $12 \times 12 \times 12$ (3456 atoms) for the point-charge spinel and zincblende lattices, respectively.
The MC simulations are performed at $10^{15}$ K.
Thermodynamic properties are evaluated using canonical MC simulations with the Metropolis algorithm\cite{metropolis}.
Thermodynamic averages are calculated for 5000 MC steps per cation after equilibration.
Even when considering a large number of MC steps, numerical errors for the thermodynamic average are small.
However, the MC simulation is not practically applicable to estimating the energy of the perfectly disordered structure by DFT calculation.

On the other hand, it is practical to use the SQS to estimate the energy of the perfectly disordered structure by DFT calculation.
SQSs are explored within a fixed size of supercell using simulated annealing instead of by computing correlation functions for all possible configurations.
In this scheme, the accuracy of the SQS is determined by the supercell size $L$ and the number of clusters $m$ used to optimize the SQS.
Since a unique solution cannot be obtained by simulated annealing, the simulated annealing is repeated ten times for each $L$ and $m$.
SQSs for the spinel and zincblende lattices are searched for using several types of supercell that are up to $6 \times 6 \times 6$  and $12 \times 12 \times 12$ expansions of the primitive cells, respectively.
Moreover, several pairs up to the 300th and 55th NNs are adopted to optimize the SQS in the point-charge spinel and zincblende lattices, respectively.
The MC simulations and SQS explorations are performed using the {\sc clupan} code.

\subsection{Correlation functions by supercell approaches}
\begin{figure}[tbp]
\begin{center}
\includegraphics[width=\linewidth]{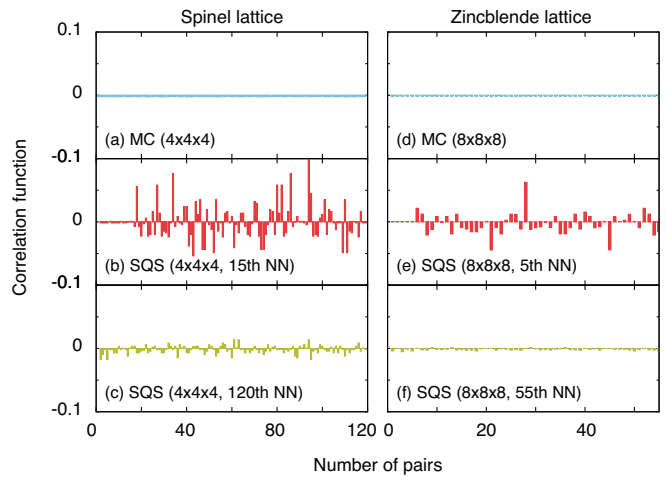} 
\caption{
(a)-(c) Differences in the correlation functions between the perfectly disordered structure and the average MC structure and between the perfectly disordered structure and an SQS with a $4 \times 4 \times 4$ supercell for the spinel lattice.
(d)-(f) Differences in the correlation functions between the perfectly disordered structure and the average MC structure and between the perfectly disordered structure and an SQS with a $8 \times 8 \times 8$ supercell for the zincblende lattice.
}
\label{sqs:sqs-correlation}
\end{center}
\end{figure}

Figure \ref{sqs:sqs-correlation} shows the differences between the average correlation functions of pairs obtained from an MC simulation at $10^{15}$ K and the correlation functions of the perfectly disordered structure for the spinel and zincblende lattices.  
The differences in the correlation functions between the SQSs with $4 \times 4 \times 4$ and $8 \times 8 \times 8$ supercells and the perfectly disordered structure for the spinel and zincblende lattices, respectively, are also shown.
For the spinel lattice, the correlation functions are optimized up to the 15th NN and 120th NN pairs, as respectively shown in Figs. \ref{sqs:sqs-correlation} (b) and (c).
For the zincblende lattice, the correlation functions are optimized up to the 5th NN and 55th NN pairs, as respectively shown in Fig. \ref{sqs:sqs-correlation} (e) and (f).
As can be seen in Fig. \ref{sqs:sqs-correlation} (a), the difference in the correlation functions between the average MC structure and the perfectly disordered structure is small and hardly any dependence on the number of pairs is observed.
On the other hand, the correlation functions of only pairs up to the 15th NN are well optimized in the SQS (15th NN) as shown in Fig. \ref{sqs:sqs-correlation} (b).
Although they are closer to those of the perfectly disordered structure than those of the average MC structure, the correlation functions of pairs longer than the 15th NN deviate considerably from those of the perfectly disordered structure.
The SQS (120th NN) also has a slightly large deviation of the correlation functions as shown in Fig. \ref{sqs:sqs-correlation} (c).
Upon increasing the supercell size, the deviation is found to be reduced.
The same tendency can be seen for the zincblende lattice.

\begin{figure}[tbp]
\begin{center}
\includegraphics[width=\linewidth]{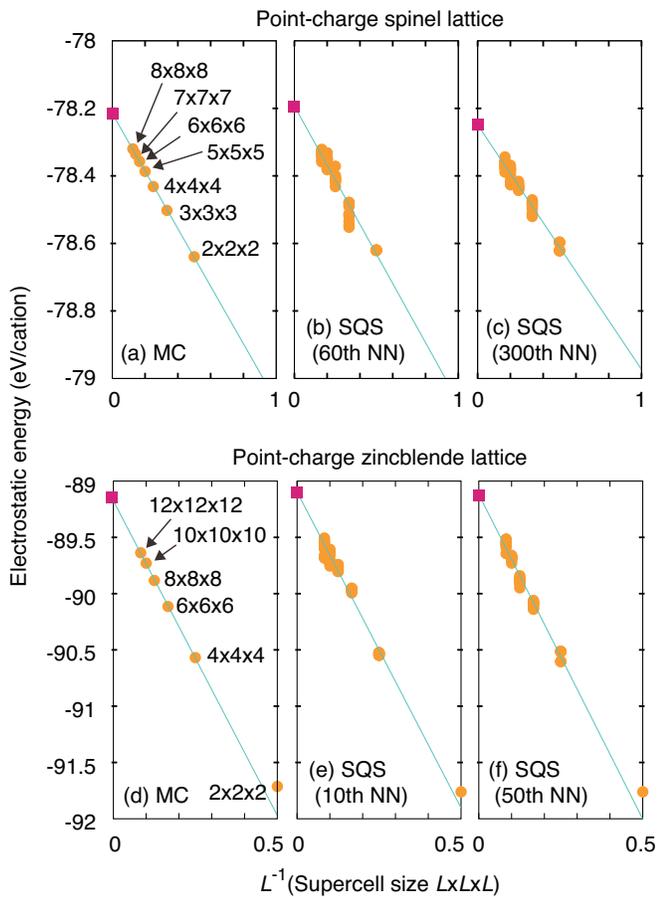} 
\caption{
Average energies obtained from MC simulations at $10^{15}$ K for (a) the point-charge spinel lattice and (d) the point-charge zincblende lattice, shown by closed circles.
The fitted line and the estimated energy of the perfectly disordered structure are shown by the solid line and closed square, respectively.
SQS energies for (b)(c) the point-charge spinel lattice and (e)(f) the point-charge zincblende lattice are also shown.
}
\label{sqs:mc-pc}
\end{center}
\end{figure}

\subsection{Average energy in MC simulation}
The deviation of the correlation functions from those of the perfectly disordered structure causes the error in the energy estimation of the perfectly disordered structure.
Figures \ref{sqs:mc-pc} (a) and (d) show the dependence of the average MC energy on the inverse of the supercell size $1/L$ for the point-charge spinel and zincblende lattices, respectively.
For the point-charge spinel lattice, even for the $L=8$ supercell (7162 atoms), the average MC energy does not converge.
Similarly, the average MC energy does not converge even for the $L=12$ supercell (3456 atoms) for the point-charge zincblende lattice.
To accurately estimate the energy of the perfectly disordered structure, extrapolation of the average MC energy to $L \to \infty$ is indispensable.
Since the average MC energy and the inverse of the supercell size are likely to have a linear relationship, the energy of the perfectly disordered structure can be estimated by extrapolation with linear regression from a set of average MC energies.  
The linear relationship is expressed as $ E_{\rm ave} = a L^{-1} + b$, where the $y$-intercept of the fitted line $b$ corresponds to the energy of the disordered structure.
The regression coefficients $a$ and $b$ are estimated using the standard least-squares technique. 
As listed in Table \ref{sqs:point-charge-disorder-energy}, the estimated energy of the disordered structure for the point-charge spinel lattice is $-78.215$ eV/cation, which is very close to the exact energy of $-78.206$ eV/cation.
In a similar manner, the estimated energy of the perfectly disordered structure for the point-charge zincblende lattice is $-89.176$ eV/cation, which is also close to the exact energy of $-89.139$ eV/cation.
These results indicate that the energy of the perfectly disordered structure can be accurately estimated from the linear extrapolation of the energies of random configurations with supercells of several sizes.

\subsection{SQS energy}
Next, the energy of the perfectly disordered structure is estimated from a set of SQS energies.
Figures \ref{sqs:mc-pc} (b) and (c) show the SQS energies obtained using pairs up to the 60th NN and 300th NN, respectively, for the point-charge spinel lattice.
Figures \ref{sqs:mc-pc} (e) and (f) show the SQS energies obtained using pairs up to the 10th NN and 50th NN, respectively, for the point-charge zincblende lattice.
Similar to the average MC energy, the SQS energy does not converge at $L=8$ and $L=12$ for the point-charge spinel and zincblende lattices, respectively.
In addition, the situation does not change upon increasing the number of pairs used to optimize the SQS, hence the selection of the number of pairs is practically less important than that of the supercell size in this case.
In contrast to the average MC energy, the SQS energies are scattered even for a fixed supercell size.
These results mean that the supercell used here is too small to find an ordered structure that can be regarded as the perfectly disordered structure.
Therefore, linear extrapolation is needed to estimate the energy of the perfectly disordered structure for both the point-charge spinel and zincblende lattices.

Linear fittings of the SQS energies are shown in Figs. \ref{sqs:mc-pc} (b), (c), (e) and (f).
The estimated energies of the disordered structure for the point-charge spinel and zincblende lattices are listed in Table \ref{sqs:point-charge-disorder-energy}.
For both the point-charge spinel and zincblende lattices, the estimated energies are close to the exact energy and slightly dependent on the choice of the number of pairs used to optimize the SQS.
To more accurately estimate the energy of the perfectly disordered structure, SQSs with larger supercells are needed.

%


\section{Applications to real systems}
So far, the energy of the perfectly disordered structure has been estimated for point-charge lattices where the exact energy of the perfectly disordered structure is known.
Next, we attempt to estimate the physical properties of the perfectly disordered structure in real systems, where the exact properties are unknown.
Here the energy and band gap of the perfectly cation-disordered structure in MgAl$_2$O$_4$ and ZnSnP$_2$ are estimated in analogy with the point-charge lattices.

DFT calculations are performed by the projector augmented-wave (PAW) method\cite{PAW1,PAW2} within the Perdew-Burke-Ernzerhof (PBE) exchange-correlation functional\cite{GGA:PBE96} as implemented in the VASP code\cite{VASP1,VASP2}.
The plane-wave cutoff energy is set to 300 eV.
The total energies converge to less than 10$^{-2}$ meV.
The atomic positions and lattice constants are relaxed until the residual forces become less than $10^{-2}$ eV$/$\AA.
DFT calculations are performed for SQSs constructed by simulated annealing using supercells. 
The similarity of the SQS and the perfectly disordered structure is defined by pairs up to the 120th and 50th NNs for MgAl$_2$O$_4$ and ZnSnP$_2$, respectively.
Supercells are constructed by isotropic expansion of the primitive lattices up to $6 \times 6 \times 6$ in both MgAl$_2$O$_4$ and ZnSnP$_2$.
Similar to the case of point-charge lattices, a unique solution cannot be obtained by simulated annealing, hence simulated annealing is repeated ten times for each supercell size.
Only in the case of a $6 \times 6 \times 6$ supercell for MgAl$_2$O$_4$ is single simulated annealing performed owing to the high computational cost.

\begin{figure}[tbp]
\begin{center}
\includegraphics[width=\linewidth]{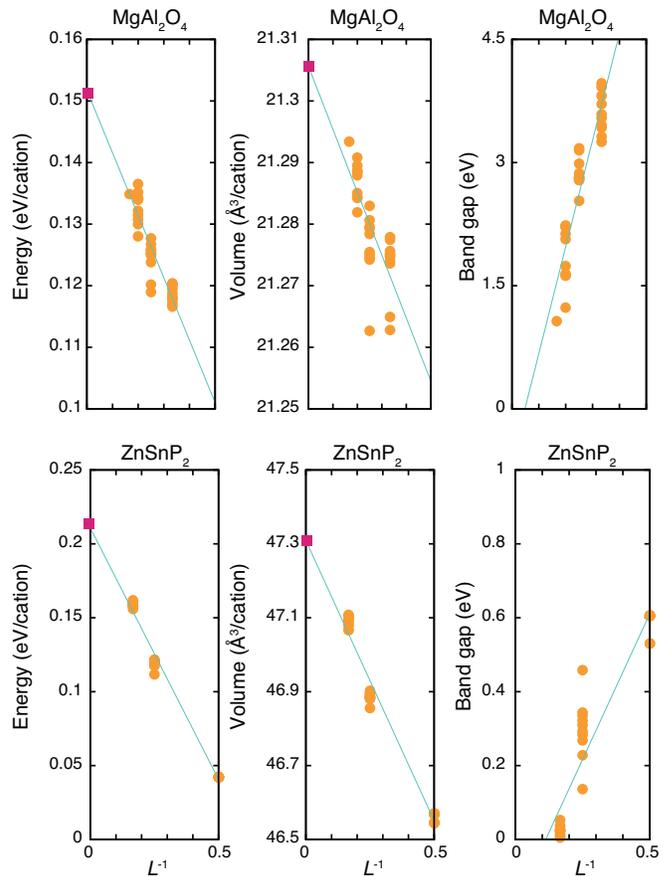} 
\caption{
Supercell size dependences of the excess energies, volumes and band gaps of the SQSs in MgAl$_2$O$_4$ and ZnSnP$_2$.
The SQSs are constructed by optimizing the correlation functions of pairs up to the 50th and 120th NNs for MgAl$_2$O$_4$ and ZnSnP$_2$, respectively.
The energies of the SQSs are measured from those of the normal spinel and chalcopyrite configurations in MgAl$_2$O$_4$ and ZnSnP$_2$, respectively.
}
\label{sqs:dft-sqs}
\end{center}
\end{figure}

Figure \ref{sqs:dft-sqs} shows the excess energies of the SQSs with several types of supercell for MgAl$_2$O$_4$ and ZnSnP$_2$, which are measured from those of the normal spinel and chalcopyrite cation configurations, respectively.
Similar to the point-charge lattices, the SQS energy does not converge with increasing supercell size.
Therefore, linear extrapolation is carried out to estimate the excess energy of the perfectly disordered structure.
The estimated excess energies of the perfectly disordered structure for MgAl$_2$O$_4$ and ZnSnP$_2$ are 0.15 and 0.21 eV/cation, which are about 0.02 and 0.05 eV/cation larger than those of the SQSs with a $6 \times 6 \times 6$ supercell, respectively.
This indicates that the energy of the perfectly disordered structure cannot be accurately estimated from a single SQS by DFT calculation with a practical computational load.
In addition to the excess energy, the volumes of the SQSs for MgAl$_2$O$_4$ and ZnSnP$_2$ are also shown in Fig. \ref{sqs:dft-sqs}.
The volume also does not converge with increasing supercell size.
By linear extrapolation of the volumes of the SQSs, the volumes of the perfectly disordered structures are estimated to be 21.306 and 47.309 \AA$^3$/cation for MgAl$_2$O$_4$ and ZnSnP$_2$, respectively.

The band gaps of the SQSs for MgAl$_2$O$_4$ and ZnSnP$_2$ are also shown in Fig. \ref{sqs:dft-sqs}.
In both MgAl$_2$O$_4$ and ZnSnP$_2$, the band gap also does not converge with increasing supercell size.
By linear extrapolation of the band gaps of the SQSs, the band gap is estimated to disappear in the perfectly disordered structure. 
In addition to the well-known underestimation of the band gap by the PBE functional, it should be noted that the band gaps estimated in this way do not necessarily correspond to those observed in experiments. 
To compare the computed band gap with the optical band gap typically used to measure the gap, one needs to compute the optical absorption spectrum and obtain the gap by fitting to an empirical equation. 
Also, Scanlon and Walsh estimated the band gap of cation-disordered ZnSnP$_2$ using a 64-atom SQS\cite{ZnSnP2_sqs_Walsh}.
Their band gap of cation-disordered ZnSnP$_2$ was 0.75 eV, obtained using the screened hybrid functional developed by Heyd, Scuseria and Ernzerhof (HSE06).
On the basis of the results of the present study, the larger band gap in their study can be ascribed not only to the use of the HSE06 functional but also to the use of a 64-atom SQS without considering the effect of the supercell size.

\section{Conclusion}
In this study, the physical properties of the perfectly disordered structure were estimated from SQS techniques for heterovalent ionic systems such as the point-charge spinel and zincblende lattices, MgAl$_2$O$_4$ and ZnSnP$_2$.
Then, their accuracy was discussed.
We found that the physical properties of the SQSs show a clear supercell size dependence and do not converge even when using a supercell considerably larger than that generally used for metallic alloys.
This originates from the fact that a large number of long-range ECIs should be considered in heterovalent ionic systems.
Therefore, to accurately estimate the physical properties of the perfectly disordered structure using the SQS, it is important to examine the convergence of the SQS properties with respect to the number of atoms.
In addition, it is found that each physical property and the inverse of the supercell size of the SQS have a linear relationship.
Therefore, the physical properties of the perfectly disordered structure can be estimated by linear extrapolation.

\begin{acknowledgments}
This study was supported by a Grant-in-Aid for Scientific Research on Innovative Areas ``Nano Informatics'' (grant number 25106005) from Japan Society for the Promotion of Science (JSPS).
\end{acknowledgments}

\bibliography{sqs}
\end{document}